\documentclass[%
 aip,
 jmp,%
 amsmath,amssymb,
reprint,%
]{revtex4-1}

\usepackage{graphicx}% Include figure files
\usepackage{dcolumn}% Align table columns on decimal point
\usepackage{bm}% bold math
\usepackage{xcolor}

\begin{document}

\title[Broadband Far-IR Absorber]{A Broadband Micro-machined Far-Infrared Absorber}
% Force line breaks with \\
%\thanks{Footnote to title of article.}
\author{E.J. Wollack}
\email{edward.j.wollack@nasa.gov}
\author{A.M. Datesman}
\author{C.A. Jhabvala}
\author{K.H. Miller}
\author{M.A. Quijada}
\affiliation{NASA Goddard Space Flight Center, Greenbelt MD 20771.
}

\date{\today}% It is always \today, today, but any date may be explicitly specified

\begin{abstract}
The experimental investigation of a broadband far-infrared meta-material absorber is described. The observed absorptance is $>\,0.95$ from ${\rm 1-20\,THz}$ (${\rm 300-15\,\mu m}$) over a temperature range spanning ${\rm 5-300\,K}$. The meta-material, realized from an array of tapers ${\rm \approx 100\,\mu m}$ in length, is largely insensitive to the detailed geometry of these elements and is cryogenically compatible with silicon-based micro-machined technologies.  The electromagnetic response is in general agreement with a physically motivated transmission line model. 
%
%Valid PACS numbers may be entered using the \verb+\pacs{#1}+ command.
\end{abstract}

%\pacs{Valid PACS appear here}% PACS, the Physics and Astronomy Classification Scheme.
\keywords{Far-Infrared, Terahertz Absorber; Meta-materials, Moth-Eye Coatings; Tapered Silicon Etch }
%Use showkeys class option if keyword
\maketitle

\section{\label{sec:introduction}Introduction}
In an ideal electromagnetic termination the incident radiation is absorbed over the desired waveband while transmission, reflection, and scattering into other propagation channels are disabled. Terminations broadly fall into two categories -- resonant and non-resonant absorbers. Resonant absorbers can be planar, compact, and conformal. The achievable bandwidth is typically narrow,~\cite{Watts2012} however, an absorptance $>\,0.98$ in over a 2:1 bandwidth has been demonstrated in a meta-material structure.~\cite{Ding2012} Non-resonant calibration targets operating in the geometric optics limit have been realized by coating conical cavities with resistive thin-films~\cite{Gush1992}  and loaded-dielectric layers~\cite{Mather1999} (absorptance $>\,0.99$ from ${\rm 10^5-100\,\mu m}$). Similarly, arrays of geometric tapers and wedges have been employed to reduce the physical envelope required for calibration targets and loads.~\cite{Janz1987,Gaidis1999,Wollack2014} Gold black,~\cite{Advena1993,Becker1999} silicon black,~\cite{Sheehy2007} and carbon nanotubes~\cite{Garcia1997,Yang2008,Quijada2011} provide examples of low-dielectric-contrast media exhibiting high-absorptance bandwidths.  

\begin{figure}
	\includegraphics[width=3.4 in]{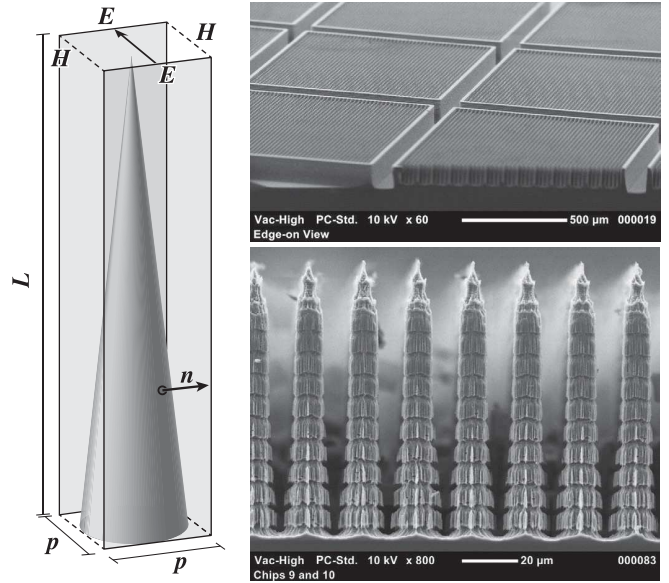}
	\vspace{0pt}
	\caption{{\it Left:} Absorber unit cell geometry and boundary conditions used in electromagnetic modeling. Here $L$ is the taper length, $p$ is the pitch of the square array lattice, and $n$ is the normal to the taper surface. Perfect electric ($E$) and magnetic ($H$) boundaries are used to represent an infinite square tiling of tapers illuminated at normal incidence. The direction of propagation $\zeta$, runs parallel to the taper length, and the polarization is indicated by an arrow near the tip of the taper.  {\it Upper right:}  A scanning electron microscope (SEM) image of the micro-machined absorber array comprised of $\approx 1 \times 1\,{\rm mm^2}$ tiles. The protective wall structures surrounding the tapers on each tile can be tailored (or omitted) to maximize the absorptance depending of the end application. The configuration depicted is compatible with the BUG (Backshort-Under-Grid) far-infrared detector architecture.~\cite{Allen2006,Miller2006} {\it Lower right:} SEM with details of the micro-machined silicon tapers on a representative absorber tile.}
	\label{fig:BackTermination}
\end{figure}

Inspired by these structures, this work explores a compact broadband absorber based on an array of micro-machined lossy silicon tapers. The approach is conceptually motivated by moth-eye anti-reflection coatings~\cite{Bernhard1967,Land2012,Southwell1991,Raguin1993} and analogous microwave terminations.~\cite{Meyer1956} Details of the meta-material's controlling parameters, fabrication, and characterization are provided in Sections~\ref{sec:theory}~\ref{sec:fabrication}, and~\ref{sec:characterization} respectively. See Figure~\ref{fig:BackTermination} for a schematic drawing of the graded-index meta-material absorber unit cell geometry ({\it left}) and images of the absorber structure ({\it right}). 

\section{\label{sec:theory}Adiabatic Absorber Theory}
The Riccatii equation provides a convenient framework to evaluate the propagation properties of an adiabatic impedance taper.~\cite{Collin1992} In its  linearized form the solution links the complex reflection amplitude, $\Gamma \ll 1 $, and the logarithmic derivative of the relative impedance, $Z_r$, along the axis of propagation, $\zeta$, 
\begin{equation}
	\Gamma =  {1 \over 2 } \cdot \int^\infty_{0} d\zeta \cdot  { e^{ {2i \beta \cdot \zeta} } \cdot {\partial \over \partial\zeta} \ln{Z_r(\zeta)}  },
	\label{equ:gamma}
\end{equation}
through a Fourier transform. At a fundamental level the achievable reflectance, bandwidth, and thickness are linked through a Kramers-Kronig relationship.~\cite{Rozanov2000} More generally, the phase constant, $\beta$, is a function of position. To the extent the fractional change in the taper's parameters is small and the conversion to other accessible modes is reversible along the structure -- the media is adiabatic. In practice, absorbing structures can be analyzed by relaxing the latter constraint.~\cite{Kuester1994,Baekelandt1999,Wollack2007} Considering the contribution from each incremental section of the taper and treating the attenuation constant, $\alpha$, as a perturbation enables a suitable generalization of Equation~\ref{equ:gamma} for the evaluation of the response in the presence of finite losses,

\begin{equation}
	\Gamma \simeq {1 \over 2 } \cdot \int^{\infty}_0 d\zeta \cdot  { e^{ { -2 \int^\zeta_0 d\zeta' \cdot \gamma(\zeta') } } \cdot {\partial \over \partial\zeta} \ln{Z_r(\zeta)}  },
	\label{equ:gamma_pert}
\end{equation}
where $\gamma = \alpha - i\beta = 2\pi i \sqrt { {\hat \mu_r} \cdot {\hat \epsilon_r}(\zeta) } /\lambda$ is the propagation constant, $Z_r(\zeta) = \sqrt{ {\hat \mu_r}    /  {\hat \epsilon_r}(\zeta) }$ is the relative wave impedance as a function of position presented by the unit cell, ${\hat \mu_r} =1$ is the relative magnetic permeability, and $\lambda$ is the freespace radiation wavelength. For $\lambda \gg p$ the effective relative dielectric function, ${\hat \epsilon_r(\zeta)}$, is quasi-static and computed from the silicon volume filling fraction for the unit cell in the mean field approximation.~\cite{Raguin1993} The dielectric losses arising from the silicon taper volume are readily treated within this framework. 

The structure's attenuation is augmented through the use of a conformal thin-film disordered-alloy coating on the micro-machined silicon surface. This allows greater control and flexibility in extinguishing incident fields. The logarithmic derivative of the impedance with respect to the normal at the taper's surface,~{\cite{Wheeler1942,Changhong2002}}
\begin{equation}
	\alpha = {R_{sq} \over 2 \eta_{\circ} } \cdot {\partial \over \partial n} \ln\left({Z_o/\eta_o}\right),
	\label{equ:loss}
\end{equation}
enables evaluation of the ohmic loss per unit length in the presence of a resistive film. The impedance of free space is $\eta_o \simeq 377 \, \Omega/\Box$ and provided the metallization layer thickness is small compared to the electrical penetration depth, $\delta$, the surface impedance, $R_{sq} \simeq \rho/t$, approximates a thin-film resistor.~\cite{Garg1975}  The boundary conditions of the perturbed system are equivalent to an infinite array of lines centered in a parallel plate guide.~\cite{ITT1975} This allows the ohmic loss arising from taper surface metallization in Equation~\ref{equ:loss} to be evaluated via,
\begin{equation}
	{Z_o(\zeta)} = { {\eta_o \over 2 \pi}  }  \cdot { \ln \left(s \cdot {p \over d(\zeta)}  \right)},
	\label{equ:coaximpedance}
\end{equation}
where $p$ is the lattice periodicity, $d(\zeta)$ is the taper diameter as a function of the length, and $s = 4/\pi$ is the inner-taper shielding factor.

With the tapered meta-material response specified, the optical properties of the structure are summarized. Incident radiation can be absorbed, specularly reflected, transmitted, or scattered to large angles. For high absorptance the taper length must be sufficient to suitably extinguish the signal traversing the structure. With increasing wavelength the meta-material surface appears smooth with respect to $\lambda$, the wave evanescently propagates into the structure and is specularly reflected before absorption. As a result, the low-frequency absorption response is limited by the total taper length (${\lambda_{\rm max} < 2L}$) for attenuation profiles consistent with the derivation of Equation~\ref{equ:gamma_pert}. 

Limitations to the high-frequency response arise from the array tiling pitch and minimum tip diameter leading to diffusive scattering into a spectrum of higher order modes (${d_{\rm tip}} \ll \lambda_{\rm min} < 2p$).~\cite{Shirley1988}  Similarly, the energy coupled to the taper is bound to the silicon volume and surface with increasing losses to radiation and internal modes as $d/\lambda$ and dielectric contrast decreases. These propagation mechanisms potentially distract from the ideal and can be mitigated through the choice of $R_{sq}$. Within the bulk silicon substrate ($\epsilon_r^{\rm Si}=11.7$), $R_{sq} = \eta_o/(\sqrt{\epsilon_r^{\rm Si}}-1) \approx 157 \, \Omega/\Box$, leads to frequency independent absorptance ($A=0.5$), reflectance ($R=0.3$), and transmittance ($T=0.2$) at normal incidence.~\cite{Carli1981} This value is adopted to specify loss in the single-mode limit and provide a graceful mechanism to extend the structure's absorption into the geometric optics regime. In the limit, $\lambda \ll 2p$, ray tracing can be used to estimate the loss at each interaction with the taper's surface.~\cite{Wolff1994} 

These behaviors suggest that a strategy to maximize the absorption bandwidth is to minimize the dielectric contrast, maximize the ratio of $L/p$, while retaining structural integrity. Within our fabrication processes, these considerations broadly lead to a lattice of tapers with suitably sharpened tip geometry for silicon's index,~\cite{Southwell1991,Alison1987} a pitch $p=20\,\mu {\rm m}$, and a length $L=100\,\mu {\rm m}$ set by the lowest wavelength of interest. 

\begin{figure}[!h]
	\includegraphics[width=3.4 in]{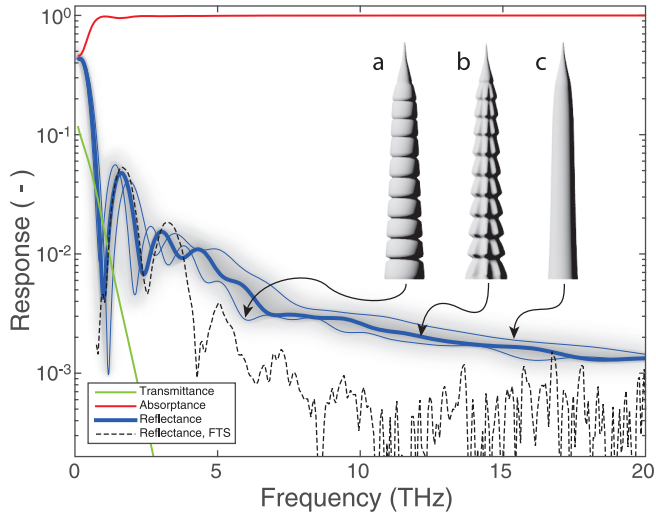}
	\vspace{0pt}
	\caption{Comparison between the observed and modeled optical response. Specular reflectance ({\it blue-solid lines}), absorptance ({\it red-solid line}), and transmittance ({\it green-solid line}) are computed at normal incidence. The modeled reflectance for three unit cell geometries is presented: (a) a profile incorporating the azimuthal slot details along absorber length derived from a digitized SEM image, (b)  an estimated 3D profile incorporating the dominate features of the structure's observed slot and ridge geometry and (c) a profile defined by the minimal taper crossection along the length. Cases ``a'' and ``c'' are taken as rotationally symmetric and are representative of upper- and lower-bounds for the volume filling fraction profile. The modeled uncertainty is schematically indicated by the {\it gray-shaded} region. The measured specular reflectance is indicated by a {\it black-dashed line}, insensitive to temperature, and diffuse scattering is subdominant.}
	\label{fig:LogResponse}
\end{figure}

To evaluate the theoretical response of the fabricated structures we adopt a variation on the synthesis theme outlined for analysis. Scanning electron micrograph (SEM) images were digitized and the resulting profiles used to defined the transfer matrix~\cite{Yeh1988} for the taper's unit cell. In evaluating $|\Gamma|^2$ a fractional change $<0.1$ was observed in increasing the number of  sub-wavelength transmission line sections from 36 to 72. The complex refractive index was estimated from the unit cell's volume filling fraction~\cite{Raguin1993} and a Drude dielectric function for the bulk silicon.~\cite{Datta2013} The surface contribution to the attenuation arising from the thin-film coating resistance dominates cold and at room temperature away from the lower band edge for the ${\rm \rho \approx 20\,\Omega-cm}$ silicon employed. High resistivity material (${\rm \rho > 1\,k\Omega-cm}$) can be used to remove the latter dispersive effect. Within and between samples, the observed absorber performance is quite repeatable; however, comparison of the modeled and observed absorber response suggest limitations in knowledge of the derived taper geometry dominate the systematic errors  in the derived transmission line parameters.  See Figure \ref{fig:LogResponse} for the modeled transmission line model response.

%Note the measured FTS data and models are for HAWC+ Sample 13D-17-10  

In the single mode limit the presence of random defects in the array induce a perturbation in the uniformity of the meta-material's surface impedance and will lead to a localized increase in observed specular reflectance. In the multi-mode limit, if defective pillars are present, the absorptance and diffuse scattering will be degraded due to a reduction in the number of bounces. From this perspective, it is desirable to achieve an areal defect fraction in the meta-material small compared to the design reflectance.~\cite{Datta2013}

Strictly speaking the meta-material structure described is a uniaxial dielectric media with propagation along the optical axis. Thus attenuance and reflectance will gracefully degrade from the perturbative treatment presented here as the polarization difference between the modeled and actual Fresnel coefficients grow with increasing angle. For incident angles $<\pi/12$ (i.e., termination of optical beams with focal ratio, $f/ > 2$, at normal incidence) the influence of this effect on the attenuance~\cite{Garcia1997,Yeh1988} is found to be subdominant to the model uncertainty imposed by the detailed knowledge of the taper geometry.

\section{\label{sec:fabrication}Sample Preparation}
The ${\rm 16\times20}$ array of square absorber tiles, ${\rm 1055\,\mu m}$ on a side, was realized via micro-machining. See Figure~\ref{fig:BackTermination} ({\it right}) for SEM images of the final absorber array structure. First, double-side-polished silicon wafers (${\rm 250\,\mu m}$ thick) were bonded to a carrier with clear wax for physical robustness during processing. Next the back side of the silicon wafer was processed to define the array's absorber tile and interconnecting structural beam geometry. These features were defined by etching silicon to a depth of ${\rm \approx 120\,\mu m}$ utilizing a standard Bosch deep reactive-ion etch (DRIE) process. To maintain etch uniformity and limit plasma loading, the textured surface on the mask is limited to 32\% of the (100\,mm diameter) silicon wafer area. The device was then released from the carrier wafer, cleaned, and wax bonded face-up to a Pyrex carrier in preparation for subsequent patterning of the meta-material structure. The use of a transparent carrier wafer and clear wax facilitates front to backside alignment. 

The square lattice of tapered pillars was realized by employing a modified Bosch DRIE process described by Roxhed et al.~\cite{Roxhed2007} This process creates tapered structures in an open field by alternating cycles of anisotropic (nearly vertical) Bosch etching with intervals of isotropic dry etching of silicon.  The narrow spacing between pillars introduces complications that do not arise when etching a tapered structure surrounded by a wide-open area. For the tightly packed square patterns employed, etching following the Roxhed prescription creates a tall pillar with a $\approx 2-3^\circ$ taper and a re-entrant profile at the top. The lithographic mask consists of ${\rm 15\,\mu m}$ wide squares on a $p={\rm 20\,\mu m}$ spacing.  Although the nearly $100:1$ photoresist selectivity is sufficient, a relatively thick ${\rm 6\,\mu m}$ layer was chosen to provided latitude in processing.  The arrays of ${\rm 100\,\mu m}$ tall pillars, each tapering to a sharp point were then realized on each tile.

The pillar points were defined by isotropically shrinking the thick photoresist layer using an in-situ oxygen plasma etch. The photoresist mask on a single pillar may reliably be shrunk from a square ${\rm 15\,\mu m}$ to ${d_{\rm tip} \approx 2\,\mu{\rm m}}$ on a side while maintaining sufficient thickness to serve as a masking layer. Subsequent DRIE of silicon under the reduced mask removes the re-entrant portion of the pillars, creating a structure with a pointed end. The alleys between wafers etch to a depth of ${\rm \approx 130\,\mu m}$ and are fully released via DRIE in the time required for the pillars to reach their final depth, ${L = 100\,\mu {\rm m}}$.  The absorber arrays were released from the temporary carrier wafer with an acetone soak, cleaned in solvents, and dried. The micro-machined arrays and an optically polished silicon witness wafer were conformally coated with ${\rm \approx 10\,nm}$ titanium nitride (${\rm TiN}$) via atomic layer deposition (ALD). 

\section{\label{sec:characterization}Optical Characterization}
The absorber's reflectance and transmittance were characterized with a Bruker IFS 125HR Fourier Transform Spectrometer (FTS) in conjunction with a liquid helium cooled bolometric detector over the spectral range of ${\rm 400-15\,\mu m}$ (${\rm 0.75-20\,THz}$).  The accessory used to measure the specular reflectance at ${\rm300\,K}$ employed a collimated beam ${\rm \approx8\,mm}$ in diameter with a $10^\circ$ angle of incidence (AOI). Masks for the sample and reference were fabricated from 60 grit sandpaper to limit the aperture in relation to the effective area of the test structure in a collimated beam.  The specular reflectance at 300, 81, and ${\rm4.9\,K}$ was also measured in a focused beam ($7^\circ$ AOI) with the masked sample mounted in an Oxford cryostat and cooled via helium exchange gas.  Upon comparing the specular reflectance spectra taken in collimated- and focused-beam configurations with the sample mounted in cryostat the difference was observed to be $<0.003$. 

In order to investigate the diffuse scattering component from the meta-material structure, the total hemispherical reflectance (THR) was characterized over the wavelength range ${\rm 20-2\,\mu m}$ at ${\rm300\,K}$. The diffuse reflectance signature was subdominant to the specular component over the spectral range accessible.  The sample's transmittance measured at ${\rm300\,K}$ was $< 2\times10^{-3}$ at wavelengths shortward of ${\rm 300\,\mu m}$ and is anticipated to increase modestly upon cooling ${\rm <30\,K}$ as free carries in the bulk silicon freeze out. An optically thick metal coating on the back-side of the absorber would effectively eliminate this perturbation on the response. For a comparison of the measured and modeled response see Figure \ref{fig:LogResponse}. For visual clarity only the dominant measured component, the spectral reflectance, is presented. 

A ${\rm 400-15\,\mu m}$ transmission spectrum of the  ${\rm TiN}$ optical witness sample at 300 K was obtained with the FTS. The surface impedance over this range was resistive with ${R_{sq}\rm =160\pm4\,\Omega/\Box}$. The residual resistance ratio of the sample on cooling from 300 to ${\rm5\,K}$ was  ${ RRR = 1.05}$.   ${\rm TiN}$ is a disordered metal-dielectric alloy (or cermet) which superconducts upon cooling below its measured transition temperature (${\rm3.5\,K}$); however, the illuminating photons ($>0.3\,{\rm THz}$) have sufficient energy to break Cooper pairs and the material effectively serves as a thin-film resistor with a surface impedance approximately equal to that of the normal state.~\cite{Tinkham1975} Thermally activated free-carriers in the bulk silicon can be rendered a subdominant absorption mechanism over the temperature and spectral ranges of interest through material selection.~\cite{Datta2013} Under intense optical illumination, the photo-induced Drude conduction in the silicon~\cite{Ding1984} can potentially influence the meta-material's long wavelength response.

\section{\label{sec:conclusion}Discussion and Conclusion}
A broadband high-absorptance meta-material structure has been realized through standard Micro-Electro Mechanical (MEMS) processes and subsequently conformally coated with a disordered thin-film TiN resistor via ALD. The absorber theory presented, based on the perturbation of boundary method, allows physical insight into parameters limiting meta-materials structure's attenuance in the single-mode limit and if desired can be used to suitably modify the response for use in other wavebands. The lower band edge of the absorption response is defined by the taper impedance and loss profiles and the reflectance is limited by the minimum taper feature size. The structure's response has been experimentally investigated and found to have $>0.95$ absorption from ${\rm 300-15\,\mu m}$ in general agreement with an adiabatic transmission line model. 

In particular, the meta-material approach explored in this work introduces loading along the taperÕs exterior surface. This conformal surface resistance layer introduces a controlled loss to the complex propagation constant and avoids limitations to the bandwidth which can arise in discretely loaded meta-material structures.~\cite{Ding2012} Without this resistive layer these structures are is equivalent to a graded index anti-reflection or ``moth-eye" coatings. The meta-material's electromagnetic topology and use of adiabatic elements is ultimately responsible for the improved absorption bandwidth observed here. 

In comparing this approach in detail to the palette of absorptive coating options for the far-infrared~\cite{Persky1999} -- namely black paints, loaded epoxy mixtures, and low-contrast lossy media -- this low-profile structure has high thickness uniformity and achieves low reflectance (${\lambda < 2L}$) through the use of a graded index. Relative to amorphous carbon and glass bead epoxy loaded coatings~\cite{Voellmer2004} this meta-material absorber structure has a lower mechanical profile, stress, and optical reflectance. Similar to black coatings produced by vapor and electrolytic deposition~\cite{Betts1985} the structure is consistent with the low-outgassing the demands of vacuum, cryogenic, and spaceflight applications.~\cite{Kauder2006,Kralik2009,Salomon2009} The observed broadband absorptance can be essentially rendered independent of temperature through material selection and the structure is fully cryogenically compatible with silicon-based micro-machined structures. 

These general properties enable the use of the meta-material structure as a broadband planar back-termination for bolometric sensor arrays~\cite{Carli1981} or as an optical glint reduction media in the ${\rm 1-20\,THz}$ spectral band. A realization of this back-termination absorber structure has been successfully implemented in the HAWC+ (High-resolution Airborne Wideband Camera) polarimeter on SOFIA (Stratospheric Observatory for Infrared Astronomy).~\cite{Jhabvala2016,Staguhn2016}

\vspace{0pt}
\section*{Acknowledgements} 
A second generation SOFIA (Stratospheric Observatory for Infrared Astronomy) instrumentation award provided by the National Aeronautics and Space Administration under NNH08ZDA009O-SOFIA2G is gratefully acknowledged by the authors.


\begin{thebibliography}{46}%
\makeatletter
\providecommand \@ifxundefined [1]{%
 \@ifx{#1\undefined}
}%
\providecommand \@ifnum [1]{%
 \ifnum #1\expandafter \@firstoftwo
 \else \expandafter \@secondoftwo
 \fi
}%
\providecommand \@ifx [1]{%
 \ifx #1\expandafter \@firstoftwo
 \else \expandafter \@secondoftwo
 \fi
}%
\providecommand \natexlab [1]{#1}%
\providecommand \enquote  [1]{``#1''}%
\providecommand \bibnamefont  [1]{#1}%
\providecommand \bibfnamefont [1]{#1}%
\providecommand \citenamefont [1]{#1}%
\providecommand \href@noop [0]{\@secondoftwo}%
\providecommand \href [0]{\begingroup \@sanitize@url \@href}%
\providecommand \@href[1]{\@@startlink{#1}\@@href}%
\providecommand \@@href[1]{\endgroup#1\@@endlink}%
\providecommand \@sanitize@url [0]{\catcode `\\12\catcode `\$12\catcode
  `\&12\catcode `\#12\catcode `\^12\catcode `\_12\catcode `\%12\relax}%
\providecommand \@@startlink[1]{}%
\providecommand \@@endlink[0]{}%
\providecommand \url  [0]{\begingroup\@sanitize@url \@url }%
\providecommand \@url [1]{\endgroup\@href {#1}{\urlprefix }}%
\providecommand \urlprefix  [0]{URL }%
\providecommand \Eprint [0]{\href }%
\providecommand \doibase [0]{http://dx.doi.org/}%
\providecommand \selectlanguage [0]{\@gobble}%
\providecommand \bibinfo  [0]{\@secondoftwo}%
\providecommand \bibfield  [0]{\@secondoftwo}%
\providecommand \translation [1]{[#1]}%
\providecommand \BibitemOpen [0]{}%
\providecommand \bibitemStop [0]{}%
\providecommand \bibitemNoStop [0]{.\EOS\space}%
\providecommand \EOS [0]{\spacefactor3000\relax}%
\providecommand \BibitemShut  [1]{\csname bibitem#1\endcsname}%
\let\auto@bib@innerbib\@empty
%</preamble>
\bibitem [{\citenamefont {Watts}, \citenamefont {Liu},\ and\ \citenamefont
  {Padilla}(2012)}]{Watts2012}%
  \BibitemOpen
  \bibfield  {author} {\bibinfo {author} {\bibfnamefont {C.~M.}\ \bibnamefont
  {Watts}}, \bibinfo {author} {\bibfnamefont {X.}~\bibnamefont {Liu}}, \ and\
  \bibinfo {author} {\bibfnamefont {W.~J.}\ \bibnamefont {Padilla}},\
  }\bibfield  {title} {\enquote {\bibinfo {title} {{Metamaterial
  Electromagnetic Wave Absorbers}},}\ }\href {\doibase 10.1002/adma.201200674}
  {\bibfield  {journal} {\bibinfo  {journal} {Advanced Materials}\ }\textbf
  {\bibinfo {volume} {24}},\ \bibinfo {pages} {OP98--OP120} (\bibinfo {year}
  {2012})}\BibitemShut {NoStop}%
\bibitem [{\citenamefont {{Ding}}\ \emph {et~al.}(2012)\citenamefont {{Ding}},
  \citenamefont {{Cui}}, \citenamefont {{Ge}}, \citenamefont {{Jin}},\ and\
  \citenamefont {{He}}}]{Ding2012}%
  \BibitemOpen
  \bibfield  {author} {\bibinfo {author} {\bibfnamefont {F.}~\bibnamefont
  {{Ding}}}, \bibinfo {author} {\bibfnamefont {Y.}~\bibnamefont {{Cui}}},
  \bibinfo {author} {\bibfnamefont {X.}~\bibnamefont {{Ge}}}, \bibinfo {author}
  {\bibfnamefont {Y.}~\bibnamefont {{Jin}}}, \ and\ \bibinfo {author}
  {\bibfnamefont {S.}~\bibnamefont {{He}}},\ }\bibfield  {title} {\enquote
  {\bibinfo {title} {{Ultra-broadband microwave metamaterial absorber}},}\
  }\href {\doibase 10.1063/1.3692178} {\bibfield  {journal} {\bibinfo
  {journal} {Applied Physics Letters}\ }\textbf {\bibinfo {volume} {100}},\
  \bibinfo {eid} {103506} (\bibinfo {year} {2012})}\BibitemShut {NoStop}%
\bibitem [{\citenamefont {Gush}\ and\ \citenamefont
  {Halpern}(1992)}]{Gush1992}%
  \BibitemOpen
  \bibfield  {author} {\bibinfo {author} {\bibfnamefont {H.~P.}\ \bibnamefont
  {Gush}}\ and\ \bibinfo {author} {\bibfnamefont {M.}~\bibnamefont {Halpern}},\
  }\bibfield  {title} {\enquote {\bibinfo {title} {{Cooled submillimeter
  Fourier transform spectrometer flown on a rocket}},}\ }\href {\doibase
  10.1063/1.1142534} {\bibfield  {journal} {\bibinfo  {journal} {Review of
  Scientific Instruments}\ }\textbf {\bibinfo {volume} {63}},\ \bibinfo {pages}
  {3249--3260} (\bibinfo {year} {1992})}\BibitemShut {NoStop}%
\bibitem [{\citenamefont {{Mather}}\ \emph {et~al.}(1999)\citenamefont
  {{Mather}}, \citenamefont {{Fixsen}}, \citenamefont {{Shafer}}, \citenamefont
  {{Mosier}},\ and\ \citenamefont {{Wilkinson}}}]{Mather1999}%
  \BibitemOpen
  \bibfield  {author} {\bibinfo {author} {\bibfnamefont {J.~C.}\ \bibnamefont
  {{Mather}}}, \bibinfo {author} {\bibfnamefont {D.~J.}\ \bibnamefont
  {{Fixsen}}}, \bibinfo {author} {\bibfnamefont {R.~A.}\ \bibnamefont
  {{Shafer}}}, \bibinfo {author} {\bibfnamefont {C.}~\bibnamefont {{Mosier}}},
  \ and\ \bibinfo {author} {\bibfnamefont {D.~T.}\ \bibnamefont
  {{Wilkinson}}},\ }\bibfield  {title} {\enquote {\bibinfo {title} {{Calibrator
  Design for the COBE Far-Infrared Absolute Spectrophotometer (FIRAS)}},}\
  }\href {\doibase 10.1086/306805} {\bibfield  {journal} {\bibinfo  {journal}
  {ApJ}\ }\textbf {\bibinfo {volume} {512}},\ \bibinfo {pages} {511--520}
  (\bibinfo {year} {1999})}\BibitemShut {NoStop}%
\bibitem [{\citenamefont {{Janz}}, \citenamefont {{Boyd}},\ and\ \citenamefont
  {{Ellis}}(1987)}]{Janz1987}%
  \BibitemOpen
  \bibfield  {author} {\bibinfo {author} {\bibfnamefont {S.}~\bibnamefont
  {{Janz}}}, \bibinfo {author} {\bibfnamefont {D.~A.}\ \bibnamefont {{Boyd}}},
  \ and\ \bibinfo {author} {\bibfnamefont {R.~F.}\ \bibnamefont {{Ellis}}},\
  }\bibfield  {title} {\enquote {\bibinfo {title} {{Reflectance characteristics
  in the submillimeter and millimeter wavelength region of a vacuum compatible
  absorber}},}\ }\href {\doibase 10.1007/BF01010634} {\bibfield  {journal}
  {\bibinfo  {journal} {International Journal of Infrared and Millimeter
  Waves}\ }\textbf {\bibinfo {volume} {8}},\ \bibinfo {pages} {627--635}
  (\bibinfo {year} {1987})}\BibitemShut {NoStop}%
\bibitem [{\citenamefont {Gaidis}, \citenamefont {Anderson},\ and\
  \citenamefont {Harding}(1999)}]{Gaidis1999}%
  \BibitemOpen
  \bibfield  {author} {\bibinfo {author} {\bibfnamefont {M.~C.}\ \bibnamefont
  {Gaidis}}, \bibinfo {author} {\bibfnamefont {M.~S.}\ \bibnamefont
  {Anderson}}, \ and\ \bibinfo {author} {\bibfnamefont {D.~G.}\ \bibnamefont
  {Harding}},\ }\bibfield  {title} {\enquote {\bibinfo {title} {{Calibration
  target for far-infrared spaceborne applications}},}\ }\href {\doibase
  10.1117/12.370182} {\bibfield  {journal} {\bibinfo  {journal} {Proc. SPIE}\
  }\textbf {\bibinfo {volume} {3795}},\ \bibinfo {pages} {348--356} (\bibinfo
  {year} {1999})}\BibitemShut {NoStop}%
\bibitem [{\citenamefont {{Wollack}}, \citenamefont {{Kinzer}},\ and\
  \citenamefont {{Rinehart}}(2014)}]{Wollack2014}%
  \BibitemOpen
  \bibfield  {author} {\bibinfo {author} {\bibfnamefont {E.~J.}\ \bibnamefont
  {{Wollack}}}, \bibinfo {author} {\bibfnamefont {R.~E.}\ \bibnamefont
  {{Kinzer}}}, \ and\ \bibinfo {author} {\bibfnamefont {S.~A.}\ \bibnamefont
  {{Rinehart}}},\ }\bibfield  {title} {\enquote {\bibinfo {title} {{A cryogenic
  infrared calibration target}},}\ }\href {\doibase 10.1063/1.4871108}
  {\bibfield  {journal} {\bibinfo  {journal} {Review of Scientific
  Instruments}\ }\textbf {\bibinfo {volume} {85}},\ \bibinfo {eid} {044707}
  (\bibinfo {year} {2014})}\BibitemShut {NoStop}%
\bibitem [{\citenamefont {Advena}, \citenamefont {Bly},\ and\ \citenamefont
  {Cox}(1993)}]{Advena1993}%
  \BibitemOpen
  \bibfield  {author} {\bibinfo {author} {\bibfnamefont {D.}~\bibnamefont
  {Advena}}, \bibinfo {author} {\bibfnamefont {V.}~\bibnamefont {Bly}}, \ and\
  \bibinfo {author} {\bibfnamefont {J.}~\bibnamefont {Cox}},\ }\bibfield
  {title} {\enquote {\bibinfo {title} {{Deposition and Characterization of
  Far-Infrared Absorbing Gold Black Films}},}\ }\href@noop {} {\bibfield
  {journal} {\bibinfo  {journal} {Appl. Opt.}\ }\textbf {\bibinfo {volume}
  {32}},\ \bibinfo {pages} {1136--1144} (\bibinfo {year} {1993})}\BibitemShut
  {NoStop}%
\bibitem [{\citenamefont {Becker}, \citenamefont {Fettig},\ and\ \citenamefont
  {Ruppel}(1999)}]{Becker1999}%
  \BibitemOpen
  \bibfield  {author} {\bibinfo {author} {\bibfnamefont {W.}~\bibnamefont
  {Becker}}, \bibinfo {author} {\bibfnamefont {R.}~\bibnamefont {Fettig}}, \
  and\ \bibinfo {author} {\bibfnamefont {W.}~\bibnamefont {Ruppel}},\
  }\bibfield  {title} {\enquote {\bibinfo {title} {Optical and electrical
  properties of black gold layers in the far infrared},}\ }\href@noop {}
  {\bibfield  {journal} {\bibinfo  {journal} {Infrared Physics \& Technology}\
  }\textbf {\bibinfo {volume} {40}},\ \bibinfo {pages} {431--445} (\bibinfo
  {year} {1999})}\BibitemShut {NoStop}%
\bibitem [{\citenamefont {Sheehy}\ \emph {et~al.}(2007)\citenamefont {Sheehy},
  \citenamefont {Tull}, \citenamefont {Friend},\ and\ \citenamefont
  {Mazur}}]{Sheehy2007}%
  \BibitemOpen
  \bibfield  {author} {\bibinfo {author} {\bibfnamefont {M.~A.}\ \bibnamefont
  {Sheehy}}, \bibinfo {author} {\bibfnamefont {B.~R.}\ \bibnamefont {Tull}},
  \bibinfo {author} {\bibfnamefont {C.~M.}\ \bibnamefont {Friend}}, \ and\
  \bibinfo {author} {\bibfnamefont {E.}~\bibnamefont {Mazur}},\ }\bibfield
  {title} {\enquote {\bibinfo {title} {{Chalcogen doping of silicon via intense
  femtosecond-laser irradiation}},}\ }\href {\doibase
  http://dx.doi.org/10.1016/j.mseb.2006.10.002} {\bibfield  {journal} {\bibinfo
   {journal} {Materials Science and Engineering: B}\ }\textbf {\bibinfo
  {volume} {137}},\ \bibinfo {pages} {289 -- 294} (\bibinfo {year}
  {2007})}\BibitemShut {NoStop}%
\bibitem [{\citenamefont {Garc\'{\i}a-Vidal}, \citenamefont {Pitarke},\ and\
  \citenamefont {Pendry}(1997)}]{Garcia1997}%
  \BibitemOpen
  \bibfield  {author} {\bibinfo {author} {\bibfnamefont {F.~J.}\ \bibnamefont
  {Garc\'{\i}a-Vidal}}, \bibinfo {author} {\bibfnamefont {J.~M.}\ \bibnamefont
  {Pitarke}}, \ and\ \bibinfo {author} {\bibfnamefont {J.~B.}\ \bibnamefont
  {Pendry}},\ }\bibfield  {title} {\enquote {\bibinfo {title} {{Effective
  Medium Theory of the Optical Properties of Aligned Carbon Nanotubes}},}\
  }\href {\doibase 10.1103/PhysRevLett.78.4289} {\bibfield  {journal} {\bibinfo
   {journal} {Phys. Rev. Lett.}\ }\textbf {\bibinfo {volume} {78}},\ \bibinfo
  {pages} {4289--4292} (\bibinfo {year} {1997})}\BibitemShut {NoStop}%
\bibitem [{\citenamefont {Yang}\ \emph {et~al.}(2008)\citenamefont {Yang},
  \citenamefont {Ci}, \citenamefont {Bur}, \citenamefont {Lin},\ and\
  \citenamefont {Ajayan}}]{Yang2008}%
  \BibitemOpen
  \bibfield  {author} {\bibinfo {author} {\bibfnamefont {Z.-P.}\ \bibnamefont
  {Yang}}, \bibinfo {author} {\bibfnamefont {L.}~\bibnamefont {Ci}}, \bibinfo
  {author} {\bibfnamefont {J.~A.}\ \bibnamefont {Bur}}, \bibinfo {author}
  {\bibfnamefont {S.-Y.}\ \bibnamefont {Lin}}, \ and\ \bibinfo {author}
  {\bibfnamefont {P.~M.}\ \bibnamefont {Ajayan}},\ }\bibfield  {title}
  {\enquote {\bibinfo {title} {{Experimental Observation of an Extremely Dark
  Material Made By a Low-Density Nanotube Array}},}\ }\href {\doibase
  10.1021/nl072369t} {\bibfield  {journal} {\bibinfo  {journal} {Nano Letters}\
  }\textbf {\bibinfo {volume} {8}},\ \bibinfo {pages} {446--451} (\bibinfo
  {year} {2008})}\BibitemShut {NoStop}%
\bibitem [{\citenamefont {{Quijada}}\ \emph {et~al.}(2011)\citenamefont
  {{Quijada}}, \citenamefont {{Hagopian}}, \citenamefont {{Getty}},
  \citenamefont {{Kinzer}},\ and\ \citenamefont {{Wollack}}}]{Quijada2011}%
  \BibitemOpen
  \bibfield  {author} {\bibinfo {author} {\bibfnamefont {M.~A.}\ \bibnamefont
  {{Quijada}}}, \bibinfo {author} {\bibfnamefont {J.~G.}\ \bibnamefont
  {{Hagopian}}}, \bibinfo {author} {\bibfnamefont {S.}~\bibnamefont {{Getty}}},
  \bibinfo {author} {\bibfnamefont {R.~E.}\ \bibnamefont {{Kinzer}}}, \ and\
  \bibinfo {author} {\bibfnamefont {E.~J.}\ \bibnamefont {{Wollack}}},\
  }\bibfield  {title} {\enquote {\bibinfo {title} {{Hemispherical reflectance
  and emittance properties of carbon nanotubes coatings at infrared
  wavelengths}},}\ }in\ \href {\doibase 10.1117/12.894601} {\emph {\bibinfo
  {booktitle} {Society of Photo-Optical Instrumentation Engineers (SPIE)
  Conference Series}}},\ \bibinfo {series} {Society of Photo-Optical
  Instrumentation Engineers (SPIE) Conference Series}, Vol.\ \bibinfo {volume}
  {8150}\ (\bibinfo {year} {2011})\BibitemShut {NoStop}%
\bibitem [{\citenamefont {Bernhard}(1967)}]{Bernhard1967}%
  \BibitemOpen
  \bibfield  {author} {\bibinfo {author} {\bibfnamefont {C.~G.}\ \bibnamefont
  {Bernhard}},\ }\bibfield  {title} {\enquote {\bibinfo {title} {Structural and
  functional adaptation in a visual system},}\ }\href@noop {} {\bibfield
  {journal} {\bibinfo  {journal} {Endeavour}\ }\textbf {\bibinfo {volume}
  {26}},\ \bibinfo {pages} {79--84} (\bibinfo {year} {1967})}\BibitemShut
  {NoStop}%
\bibitem [{\citenamefont {Land}\ and\ \citenamefont {Nisson}(2012)}]{Land2012}%
  \BibitemOpen
  \bibfield  {author} {\bibinfo {author} {\bibfnamefont {M.~F.}\ \bibnamefont
  {Land}}\ and\ \bibinfo {author} {\bibfnamefont {D.~E.}\ \bibnamefont
  {Nisson}},\ }\href@noop {} {\emph {\bibinfo {title} {{Animal Eyes}}}},\
  \bibinfo {edition} {2nd}\ ed.\ (\bibinfo  {publisher} {Oxford Animal Biology
  Series, Oxford University Press},\ \bibinfo {year} {2012})\ p.\ \bibinfo
  {pages} {142}\BibitemShut {NoStop}%
\bibitem [{\citenamefont {Southwell}(1991)}]{Southwell1991}%
  \BibitemOpen
  \bibfield  {author} {\bibinfo {author} {\bibfnamefont {W.~H.}\ \bibnamefont
  {Southwell}},\ }\bibfield  {title} {\enquote {\bibinfo {title}
  {{Pyramid-array surface-relief structures producing antireflection index
  matching on optical surfaces}},}\ }\href {\doibase 10.1364/JOSAA.8.000549}
  {\bibfield  {journal} {\bibinfo  {journal} {J. Opt. Soc. Am. A}\ }\textbf
  {\bibinfo {volume} {8}},\ \bibinfo {pages} {549--553} (\bibinfo {year}
  {1991})}\BibitemShut {NoStop}%
\bibitem [{\citenamefont {Raguin}\ and\ \citenamefont
  {Morris}(1993)}]{Raguin1993}%
  \BibitemOpen
  \bibfield  {author} {\bibinfo {author} {\bibfnamefont {D.~H.}\ \bibnamefont
  {Raguin}}\ and\ \bibinfo {author} {\bibfnamefont {G.~M.}\ \bibnamefont
  {Morris}},\ }\bibfield  {title} {\enquote {\bibinfo {title} {{Antireflection
  structured surfaces for the infrared spectral region}},}\ }\href {\doibase
  10.1364/AO.32.001154} {\bibfield  {journal} {\bibinfo  {journal} {Appl.
  Opt.}\ }\textbf {\bibinfo {volume} {32}},\ \bibinfo {pages} {1154--1167}
  (\bibinfo {year} {1993})}\BibitemShut {NoStop}%
\bibitem [{\citenamefont {Meyer}\ and\ \citenamefont
  {Severin}(1956)}]{Meyer1956}%
  \BibitemOpen
  \bibfield  {author} {\bibinfo {author} {\bibfnamefont {E.}~\bibnamefont
  {Meyer}}\ and\ \bibinfo {author} {\bibfnamefont {H.}~\bibnamefont
  {Severin}},\ }\bibfield  {title} {\enquote {\bibinfo {title} {{Absorption
  devices for electromagnetic centimeter waves and their acoustic analogs}},}\
  }\href@noop {} {\bibfield  {journal} {\bibinfo  {journal} {Z. Angew. Phys.}\
  }\textbf {\bibinfo {volume} {8}},\ \bibinfo {pages} {105 --114} (\bibinfo
  {year} {1956})}\BibitemShut {NoStop}%
\bibitem [{\citenamefont {{Allen}}\ \emph {et~al.}(2006)\citenamefont
  {{Allen}}, \citenamefont {{Benford}}, \citenamefont {{Chervenak}},
  \citenamefont {{Chuss}}, \citenamefont {{Miller}}, \citenamefont {{Moseley}},
  \citenamefont {{Staguhn}},\ and\ \citenamefont {J.{Wollack}}}]{Allen2006}%
  \BibitemOpen
  \bibfield  {author} {\bibinfo {author} {\bibfnamefont {C.~A.}\ \bibnamefont
  {{Allen}}}, \bibinfo {author} {\bibfnamefont {D.~J.}\ \bibnamefont
  {{Benford}}}, \bibinfo {author} {\bibfnamefont {J.~A.}\ \bibnamefont
  {{Chervenak}}}, \bibinfo {author} {\bibfnamefont {D.~T.}\ \bibnamefont
  {{Chuss}}}, \bibinfo {author} {\bibfnamefont {T.~M.}\ \bibnamefont
  {{Miller}}}, \bibinfo {author} {\bibfnamefont {S.~H.}\ \bibnamefont
  {{Moseley}}}, \bibinfo {author} {\bibfnamefont {J.~G.}\ \bibnamefont
  {{Staguhn}}}, \ and\ \bibinfo {author} {\bibfnamefont {E.}~\bibnamefont
  {J.{Wollack}}},\ }\bibfield  {title} {\enquote {\bibinfo {title}
  {{Backshort-Under-Grid arrays for infrared astronomy}},}\ }\href {\doibase
  10.1016/j.nima.2005.12.059} {\bibfield  {journal} {\bibinfo  {journal}
  {Nuclear Instruments and Methods in Physics Research A}\ }\textbf {\bibinfo
  {volume} {559}},\ \bibinfo {pages} {522--524} (\bibinfo {year}
  {2006})}\BibitemShut {NoStop}%
\bibitem [{\citenamefont {Miller}, \citenamefont {Abrahams},\ and\
  \citenamefont {Allen}(2006)}]{Miller2006}%
  \BibitemOpen
  \bibfield  {author} {\bibinfo {author} {\bibfnamefont {T.~M.}\ \bibnamefont
  {Miller}}, \bibinfo {author} {\bibfnamefont {J.~H.}\ \bibnamefont
  {Abrahams}}, \ and\ \bibinfo {author} {\bibfnamefont {C.~A.}\ \bibnamefont
  {Allen}},\ }\bibfield  {title} {\enquote {\bibinfo {title} {{Fabricating
  interlocking support walls, with an adjustable backshort, in a TES bolometer
  array for far-infrared astronomy}},}\ }\href {\doibase
  http://dx.doi.org/10.1016/j.nima.2005.12.067} {\bibfield  {journal} {\bibinfo
   {journal} {Nuclear Instruments and Methods in Physics Research Section A}\
  }\textbf {\bibinfo {volume} {559}},\ \bibinfo {pages} {548 -- 550} (\bibinfo
  {year} {2006})},\ \bibinfo {note} {proceedings of the 11th International
  Workshop on Low Temperature Detectors}\BibitemShut {NoStop}%
\bibitem [{\citenamefont {Collin}(1992)}]{Collin1992}%
  \BibitemOpen
  \bibfield  {author} {\bibinfo {author} {\bibfnamefont {R.~E.}\ \bibnamefont
  {Collin}},\ }\href@noop {} {\emph {\bibinfo {title} {{Foundations for
  Microwave Engineering}}}},\ \bibinfo {edition} {2nd}\ ed.\ (\bibinfo
  {publisher} {McGraw-Hill},\ \bibinfo {address} {New York},\ \bibinfo {year}
  {1992})\BibitemShut {NoStop}%
\bibitem [{\citenamefont {Rozanov}(2000)}]{Rozanov2000}%
  \BibitemOpen
  \bibfield  {author} {\bibinfo {author} {\bibfnamefont {K.}~\bibnamefont
  {Rozanov}},\ }\bibfield  {title} {\enquote {\bibinfo {title} {{Ultimate
  thickness to bandwidth ratio of radar absorbers}},}\ }\href {\doibase
  10.1109/8.884491} {\bibfield  {journal} {\bibinfo  {journal} {Antennas and
  Propagation, IEEE Transactions on}\ }\textbf {\bibinfo {volume} {48}},\
  \bibinfo {pages} {1230--1234} (\bibinfo {year} {2000})}\BibitemShut {NoStop}%
\bibitem [{\citenamefont {Kuester}\ and\ \citenamefont
  {Holloway}(1994)}]{Kuester1994}%
  \BibitemOpen
  \bibfield  {author} {\bibinfo {author} {\bibfnamefont {E.~F.}\ \bibnamefont
  {Kuester}}\ and\ \bibinfo {author} {\bibfnamefont {C.}~\bibnamefont
  {Holloway}},\ }\bibfield  {title} {\enquote {\bibinfo {title} {{A
  low-frequency model for wedge or pyramid absorber arrays-I: theory}},}\
  }\href {\doibase 10.1109/15.328859} {\bibfield  {journal} {\bibinfo
  {journal} {Electromagnetic Compatibility, IEEE Transactions on}\ }\textbf
  {\bibinfo {volume} {36}},\ \bibinfo {pages} {300--306} (\bibinfo {year}
  {1994})}\BibitemShut {NoStop}%
\bibitem [{\citenamefont {Baekelandt}, \citenamefont {Olyslager},\ and\
  \citenamefont {De~Zutter}(1999)}]{Baekelandt1999}%
  \BibitemOpen
  \bibfield  {author} {\bibinfo {author} {\bibfnamefont {B.}~\bibnamefont
  {Baekelandt}}, \bibinfo {author} {\bibfnamefont {F.}~\bibnamefont
  {Olyslager}}, \ and\ \bibinfo {author} {\bibfnamefont {D.}~\bibnamefont
  {De~Zutter}},\ }\bibfield  {title} {\enquote {\bibinfo {title}
  {{Low-frequency reflectivity approximation for two- and three-dimensional EM
  absorbers}},}\ }\href {\doibase 10.1109/15.809815} {\bibfield  {journal}
  {\bibinfo  {journal} {Electromagnetic Compatibility, IEEE Transactions on}\
  }\textbf {\bibinfo {volume} {41}},\ \bibinfo {pages} {354--360} (\bibinfo
  {year} {1999})}\BibitemShut {NoStop}%
\bibitem [{\citenamefont {Wollack}\ \emph {et~al.}(2007)\citenamefont
  {Wollack}, \citenamefont {Fixsen}, \citenamefont {Kogut}, \citenamefont
  {Limon}, \citenamefont {Mirel},\ and\ \citenamefont {Singal}}]{Wollack2007}%
  \BibitemOpen
  \bibfield  {author} {\bibinfo {author} {\bibfnamefont {E.}~\bibnamefont
  {Wollack}}, \bibinfo {author} {\bibfnamefont {D.}~\bibnamefont {Fixsen}},
  \bibinfo {author} {\bibfnamefont {A.}~\bibnamefont {Kogut}}, \bibinfo
  {author} {\bibfnamefont {M.}~\bibnamefont {Limon}}, \bibinfo {author}
  {\bibfnamefont {P.}~\bibnamefont {Mirel}}, \ and\ \bibinfo {author}
  {\bibfnamefont {J.}~\bibnamefont {Singal}},\ }\bibfield  {title} {\enquote
  {\bibinfo {title} {{Radiometric-Waveguide Calibrators}},}\ }\href@noop {}
  {\bibfield  {journal} {\bibinfo  {journal} {IEEE T. Instrumentation and
  Measurement}\ }\textbf {\bibinfo {volume} {56}},\ \bibinfo {pages}
  {2073--2078} (\bibinfo {year} {2007})}\BibitemShut {NoStop}%
\bibitem [{\citenamefont {Wheeler}(1942)}]{Wheeler1942}%
  \BibitemOpen
  \bibfield  {author} {\bibinfo {author} {\bibfnamefont {H.~A.}\ \bibnamefont
  {Wheeler}},\ }\bibfield  {title} {\enquote {\bibinfo {title} {{Formulas for
  the Skin Effect}},}\ }\href {\doibase 10.1109/JRPROC.1942.232015} {\bibfield
  {journal} {\bibinfo  {journal} {Proceedings of the IRE}\ }\textbf {\bibinfo
  {volume} {30}},\ \bibinfo {pages} {412--424} (\bibinfo {year}
  {1942})}\BibitemShut {NoStop}%
\bibitem [{\citenamefont {Changhong}\ and\ \citenamefont
  {Long}(2002)}]{Changhong2002}%
  \BibitemOpen
  \bibfield  {author} {\bibinfo {author} {\bibfnamefont {L.}~\bibnamefont
  {Changhong}}\ and\ \bibinfo {author} {\bibfnamefont {L.}~\bibnamefont
  {Long}},\ }\bibfield  {title} {\enquote {\bibinfo {title} {{A new
  characteristic impedance perturbation method for finding attenuation
  constants}},}\ }\href {\doibase 10.1002/mop.10143} {\bibfield  {journal}
  {\bibinfo  {journal} {Microwave and Optical Technology Letters}\ }\textbf
  {\bibinfo {volume} {32}},\ \bibinfo {pages} {243--245} (\bibinfo {year}
  {2002})}\BibitemShut {NoStop}%
\bibitem [{\citenamefont {Garg}, \citenamefont {Gupta},\ and\ \citenamefont
  {Sharan}(1975)}]{Garg1975}%
  \BibitemOpen
  \bibfield  {author} {\bibinfo {author} {\bibfnamefont {R.}~\bibnamefont
  {Garg}}, \bibinfo {author} {\bibfnamefont {K.~C.}\ \bibnamefont {Gupta}}, \
  and\ \bibinfo {author} {\bibfnamefont {R.}~\bibnamefont {Sharan}},\
  }\bibfield  {title} {\enquote {\bibinfo {title} {{Optimum thickness of metal
  in waveguiding structures, ground planes and reflectors}},}\ }\href {\doibase
  10.1080/00207217508920513} {\bibfield  {journal} {\bibinfo  {journal}
  {International Journal of Electronics}\ }\textbf {\bibinfo {volume} {39}},\
  \bibinfo {pages} {525--527} (\bibinfo {year} {1975})}\BibitemShut {NoStop}%
\bibitem [{\citenamefont {ITT}(1975)}]{ITT1975}%
  \BibitemOpen
  \bibfield  {author} {\bibinfo {author} {\bibnamefont {ITT}},\ }\href@noop {}
  {\emph {\bibinfo {title} {{Reference Data for Radio Engineers}}}}\ (\bibinfo
  {publisher} {Howard W. Sams},\ \bibinfo {address} {New York},\ \bibinfo
  {year} {1975})\ Chap.~\bibinfo {chapter} {24}, pp.\ \bibinfo {pages} {21 --
  23}\BibitemShut {NoStop}%
\bibitem [{\citenamefont {Shirley}\ and\ \citenamefont
  {George}(1988)}]{Shirley1988}%
  \BibitemOpen
  \bibfield  {author} {\bibinfo {author} {\bibfnamefont {L.~G.}\ \bibnamefont
  {Shirley}}\ and\ \bibinfo {author} {\bibfnamefont {N.}~\bibnamefont
  {George}},\ }\bibfield  {title} {\enquote {\bibinfo {title} {Diffuser
  radiation patterns over a large dynamic range. 1: Strong diffusers},}\ }\href
  {\doibase 10.1364/AO.27.001850} {\bibfield  {journal} {\bibinfo  {journal}
  {Appl. Opt.}\ }\textbf {\bibinfo {volume} {27}},\ \bibinfo {pages}
  {1850--1861} (\bibinfo {year} {1988})}\BibitemShut {NoStop}%
\bibitem [{\citenamefont {Carli}\ and\ \citenamefont
  {Iorio-Fili}(1981)}]{Carli1981}%
  \BibitemOpen
  \bibfield  {author} {\bibinfo {author} {\bibfnamefont {B.}~\bibnamefont
  {Carli}}\ and\ \bibinfo {author} {\bibfnamefont {D.}~\bibnamefont
  {Iorio-Fili}},\ }\bibfield  {title} {\enquote {\bibinfo {title} {{Absorption
  of composite bolometers}},}\ }\href {\doibase 10.1364/JOSA.71.001020}
  {\bibfield  {journal} {\bibinfo  {journal} {J. Opt. Soc. Am.}\ }\textbf
  {\bibinfo {volume} {71}},\ \bibinfo {pages} {1020--1025} (\bibinfo {year}
  {1981})}\BibitemShut {NoStop}%
\bibitem [{\citenamefont {Wolff}(1994)}]{Wolff1994}%
  \BibitemOpen
  \bibfield  {author} {\bibinfo {author} {\bibfnamefont {L.~B.}\ \bibnamefont
  {Wolff}},\ }\bibfield  {title} {\enquote {\bibinfo {title}
  {Diffuse-reflectance model for smooth dielectric surfaces},}\ }\href
  {\doibase 10.1364/JOSAA.11.002956} {\bibfield  {journal} {\bibinfo  {journal}
  {J. Opt. Soc. Am. A}\ }\textbf {\bibinfo {volume} {11}},\ \bibinfo {pages}
  {2956--2968} (\bibinfo {year} {1994})}\BibitemShut {NoStop}%
\bibitem [{\citenamefont {Alison}(1987)}]{Alison1987}%
  \BibitemOpen
  \bibfield  {author} {\bibinfo {author} {\bibfnamefont {W.~B.~W.}\
  \bibnamefont {Alison}},\ }\href@noop {} {\emph {\bibinfo {title} {{A Handbook
  for the Mechnical Tolerancing of Waveguide Components}}}}\ (\bibinfo
  {publisher} {Artech House},\ \bibinfo {year} {1987})\ pp.\ \bibinfo {pages}
  {364--368}\BibitemShut {NoStop}%
\bibitem [{\citenamefont {Yeh}(1988)}]{Yeh1988}%
  \BibitemOpen
  \bibfield  {author} {\bibinfo {author} {\bibfnamefont {P.}~\bibnamefont
  {Yeh}},\ }\href@noop {} {\emph {\bibinfo {title} {{Optical Waves in Layered
  Media}}}}\ (\bibinfo  {publisher} {John Wiley \& Sons},\ \bibinfo {address}
  {New York},\ \bibinfo {year} {1988})\ Chap.~\bibinfo {chapter} {5}, pp.\
  \bibinfo {pages} {102--117}\BibitemShut {NoStop}%
\bibitem [{\citenamefont {Datta}\ \emph {et~al.}(2013)\citenamefont {Datta},
  \citenamefont {Munson}, \citenamefont {Niemack}, \citenamefont {McMahon},
  \citenamefont {Britton}, \citenamefont {Wollack}, \citenamefont {Beall},
  \citenamefont {Devlin}, \citenamefont {Fowler}, \citenamefont {Gallardo},
  \citenamefont {Hubmayr}, \citenamefont {Irwin}, \citenamefont {Newburgh},
  \citenamefont {Nibarger}, \citenamefont {Page}, \citenamefont {Quijada},
  \citenamefont {Schmitt}, \citenamefont {Staggs}, \citenamefont {Thornton},\
  and\ \citenamefont {Zhang}}]{Datta2013}%
  \BibitemOpen
  \bibfield  {author} {\bibinfo {author} {\bibfnamefont {R.}~\bibnamefont
  {Datta}}, \bibinfo {author} {\bibfnamefont {C.~D.}\ \bibnamefont {Munson}},
  \bibinfo {author} {\bibfnamefont {M.~D.}\ \bibnamefont {Niemack}}, \bibinfo
  {author} {\bibfnamefont {J.~J.}\ \bibnamefont {McMahon}}, \bibinfo {author}
  {\bibfnamefont {J.}~\bibnamefont {Britton}}, \bibinfo {author} {\bibfnamefont
  {E.~J.}\ \bibnamefont {Wollack}}, \bibinfo {author} {\bibfnamefont
  {J.}~\bibnamefont {Beall}}, \bibinfo {author} {\bibfnamefont {M.~J.}\
  \bibnamefont {Devlin}}, \bibinfo {author} {\bibfnamefont {J.}~\bibnamefont
  {Fowler}}, \bibinfo {author} {\bibfnamefont {P.}~\bibnamefont {Gallardo}},
  \bibinfo {author} {\bibfnamefont {J.}~\bibnamefont {Hubmayr}}, \bibinfo
  {author} {\bibfnamefont {K.}~\bibnamefont {Irwin}}, \bibinfo {author}
  {\bibfnamefont {L.}~\bibnamefont {Newburgh}}, \bibinfo {author}
  {\bibfnamefont {J.~P.}\ \bibnamefont {Nibarger}}, \bibinfo {author}
  {\bibfnamefont {L.}~\bibnamefont {Page}}, \bibinfo {author} {\bibfnamefont
  {M.~A.}\ \bibnamefont {Quijada}}, \bibinfo {author} {\bibfnamefont {B.~L.}\
  \bibnamefont {Schmitt}}, \bibinfo {author} {\bibfnamefont {S.~T.}\
  \bibnamefont {Staggs}}, \bibinfo {author} {\bibfnamefont {R.}~\bibnamefont
  {Thornton}}, \ and\ \bibinfo {author} {\bibfnamefont {L.}~\bibnamefont
  {Zhang}},\ }\bibfield  {title} {\enquote {\bibinfo {title} {{Large-aperture
  wide-bandwidth antireflection-coated silicon lenses for millimeter
  wavelengths}},}\ }\href {\doibase 10.1364/AO.52.008747} {\bibfield  {journal}
  {\bibinfo  {journal} {Appl. Opt.}\ }\textbf {\bibinfo {volume} {52}},\
  \bibinfo {pages} {8747--8758} (\bibinfo {year} {2013})}\BibitemShut {NoStop}%
\bibitem [{\citenamefont {Roxhed}, \citenamefont {Griss},\ and\ \citenamefont
  {Stemme}(2007)}]{Roxhed2007}%
  \BibitemOpen
  \bibfield  {author} {\bibinfo {author} {\bibfnamefont {N.}~\bibnamefont
  {Roxhed}}, \bibinfo {author} {\bibfnamefont {P.}~\bibnamefont {Griss}}, \
  and\ \bibinfo {author} {\bibfnamefont {G.}~\bibnamefont {Stemme}},\
  }\bibfield  {title} {\enquote {\bibinfo {title} {{A method for tapered deep
  reactive ion etching using a modified Bosch process}},}\ }\href
  {http://stacks.iop.org/0960-1317/17/i=5/a=031} {\bibfield  {journal}
  {\bibinfo  {journal} {Journal of Micromechanics and Microengineering}\
  }\textbf {\bibinfo {volume} {17}},\ \bibinfo {pages} {1087} (\bibinfo {year}
  {2007})}\BibitemShut {NoStop}%
\bibitem [{\citenamefont {Tinkham}(1996)}]{Tinkham1975}%
  \BibitemOpen
  \bibfield  {author} {\bibinfo {author} {\bibfnamefont {M.}~\bibnamefont
  {Tinkham}},\ }\href@noop {} {\emph {\bibinfo {title} {{Introduction to
  Superconductivty}}}}\ (\bibinfo  {publisher} {McGraw-Hill},\ \bibinfo {year}
  {1996})\ pp.\ \bibinfo {pages} {68--71}\BibitemShut {NoStop}%
\bibitem [{\citenamefont {Ding}\ \emph {et~al.}(1984)\citenamefont {Ding},
  \citenamefont {Shih}, \citenamefont {Pavlasek},\ and\ \citenamefont
  {Champness}}]{Ding1984}%
  \BibitemOpen
  \bibfield  {author} {\bibinfo {author} {\bibfnamefont {L.}~\bibnamefont
  {Ding}}, \bibinfo {author} {\bibfnamefont {I.}~\bibnamefont {Shih}}, \bibinfo
  {author} {\bibfnamefont {T.~J.}\ \bibnamefont {Pavlasek}}, \ and\ \bibinfo
  {author} {\bibfnamefont {C.}~\bibnamefont {Champness}},\ }\bibfield  {title}
  {\enquote {\bibinfo {title} {{Measurements of the Photo-Induced Complex
  Permittivity of Si, Ge, and Te at 9 GHz}},}\ }\href {\doibase
  10.1109/TMTT.1984.1132635} {\bibfield  {journal} {\bibinfo  {journal}
  {Microwave Theory and Techniques, IEEE Transactions on}\ }\textbf {\bibinfo
  {volume} {32}},\ \bibinfo {pages} {151--157} (\bibinfo {year}
  {1984})}\BibitemShut {NoStop}%
\bibitem [{\citenamefont {Persky}(1999)}]{Persky1999}%
  \BibitemOpen
  \bibfield  {author} {\bibinfo {author} {\bibfnamefont {M.}~\bibnamefont
  {Persky}},\ }\bibfield  {title} {\enquote {\bibinfo {title} {Review of black
  surfaces for space-borne infrared systems},}\ }\href {\doibase
  10.1063/1.1149739} {\bibfield  {journal} {\bibinfo  {journal} {Review of
  Scientific Instruments}\ }\textbf {\bibinfo {volume} {70}},\ \bibinfo {pages}
  {2193--2217} (\bibinfo {year} {1999})}\BibitemShut {NoStop}%
\bibitem [{\citenamefont {{Voellmer}}\ \emph {et~al.}(2004)\citenamefont
  {{Voellmer}}, \citenamefont {{Allen}}, \citenamefont {{Babu}}, \citenamefont
  {{Bartels}}, \citenamefont {{Dowell}}, \citenamefont {{Dotson}},
  \citenamefont {{Harper}}, \citenamefont {{Moseley}}, \citenamefont
  {{Rennick}}, \citenamefont {{Shirron}}, \citenamefont {{Smith}},\ and\
  \citenamefont {{Wollack}}}]{Voellmer2004}%
  \BibitemOpen
  \bibfield  {author} {\bibinfo {author} {\bibfnamefont {G.~M.}\ \bibnamefont
  {{Voellmer}}}, \bibinfo {author} {\bibfnamefont {C.~A.}\ \bibnamefont
  {{Allen}}}, \bibinfo {author} {\bibfnamefont {S.~R.}\ \bibnamefont {{Babu}}},
  \bibinfo {author} {\bibfnamefont {A.~E.}\ \bibnamefont {{Bartels}}}, \bibinfo
  {author} {\bibfnamefont {C.~D.}\ \bibnamefont {{Dowell}}}, \bibinfo {author}
  {\bibfnamefont {J.~L.}\ \bibnamefont {{Dotson}}}, \bibinfo {author}
  {\bibfnamefont {D.~A.}\ \bibnamefont {{Harper}}}, \bibinfo {author}
  {\bibfnamefont {S.~H.}\ \bibnamefont {{Moseley}}, \bibfnamefont {Jr.}},
  \bibinfo {author} {\bibfnamefont {T.}~\bibnamefont {{Rennick}}}, \bibinfo
  {author} {\bibfnamefont {P.~J.}\ \bibnamefont {{Shirron}}}, \bibinfo {author}
  {\bibfnamefont {W.~W.}\ \bibnamefont {{Smith}}}, \ and\ \bibinfo {author}
  {\bibfnamefont {E.~J.}\ \bibnamefont {{Wollack}}},\ }\bibfield  {title}
  {\enquote {\bibinfo {title} {{A two-dimensional semiconducting bolometer
  array for HAWC}},}\ \ }(\bibinfo {year} {2004})\ pp.\ \bibinfo {pages}
  {428--437}\BibitemShut {NoStop}%
\bibitem [{\citenamefont {Betts}\ \emph {et~al.}(1985)\citenamefont {Betts},
  \citenamefont {Clarke}, \citenamefont {Cox},\ and\ \citenamefont
  {Larkin}}]{Betts1985}%
  \BibitemOpen
  \bibfield  {author} {\bibinfo {author} {\bibfnamefont {D.~B.}\ \bibnamefont
  {Betts}}, \bibinfo {author} {\bibfnamefont {F.~J.~J.}\ \bibnamefont
  {Clarke}}, \bibinfo {author} {\bibfnamefont {L.~J.}\ \bibnamefont {Cox}}, \
  and\ \bibinfo {author} {\bibfnamefont {J.~A.}\ \bibnamefont {Larkin}},\
  }\bibfield  {title} {\enquote {\bibinfo {title} {Infrared reflection
  properties of five types of black coating for radiometric detectors},}\
  }\href {http://stacks.iop.org/0022-3735/18/i=8/a=010} {\bibfield  {journal}
  {\bibinfo  {journal} {Journal of Physics E: Scientific Instruments}\ }\textbf
  {\bibinfo {volume} {18}},\ \bibinfo {pages} {689--696} (\bibinfo {year}
  {1985})}\BibitemShut {NoStop}%
\bibitem [{\citenamefont {Kauder}(2006)}]{Kauder2006}%
  \BibitemOpen
  \bibfield  {author} {\bibinfo {author} {\bibfnamefont {L.}~\bibnamefont
  {Kauder}},\ }\href@noop {} {\enquote {\bibinfo {title} {Spacecraft thermal
  control coatings references},}\ }\bibinfo {type} {Tech. Rep.}\ \bibinfo
  {number} {TP-2005-212792}\ (\bibinfo  {institution} {NASA Goddard Space
  Flight Center},\ \bibinfo {year} {2006})\BibitemShut {NoStop}%
\bibitem [{\citenamefont {Kralik}\ and\ \citenamefont
  {Katsir}(2009)}]{Kralik2009}%
  \BibitemOpen
  \bibfield  {author} {\bibinfo {author} {\bibfnamefont {T.}~\bibnamefont
  {Kralik}}\ and\ \bibinfo {author} {\bibfnamefont {D.}~\bibnamefont
  {Katsir}},\ }\bibfield  {title} {\enquote {\bibinfo {title} {{Black surfaces
  for infrared, aerospace, and cryogenic applications}},}\ }in\ \href {\doibase
  10.1117/12.819277} {\emph {\bibinfo {booktitle} {Society of Photo-Optical
  Instrumentation Engineers (SPIE) Conference Series}}},\ \bibinfo {series}
  {Society of Photo-Optical Instrumentation Engineers (SPIE) Conference
  Series}, Vol.\ \bibinfo {volume} {7298}\ (\bibinfo {year} {2009})\BibitemShut
  {NoStop}%
\bibitem [{\citenamefont {Salomon}\ \emph {et~al.}(2009)\citenamefont
  {Salomon}, \citenamefont {Sternberg}, \citenamefont {Gouzman}, \citenamefont
  {Lempert}, \citenamefont {Grossman}, \citenamefont {Katsir}, \citenamefont
  {Cotostiano},\ and\ \citenamefont {Minton}}]{Salomon2009}%
  \BibitemOpen
  \bibfield  {author} {\bibinfo {author} {\bibfnamefont {Y.}~\bibnamefont
  {Salomon}}, \bibinfo {author} {\bibfnamefont {N.}~\bibnamefont {Sternberg}},
  \bibinfo {author} {\bibfnamefont {I.}~\bibnamefont {Gouzman}}, \bibinfo
  {author} {\bibfnamefont {G.}~\bibnamefont {Lempert}}, \bibinfo {author}
  {\bibfnamefont {E.}~\bibnamefont {Grossman}}, \bibinfo {author}
  {\bibfnamefont {D.}~\bibnamefont {Katsir}}, \bibinfo {author} {\bibfnamefont
  {R.}~\bibnamefont {Cotostiano}}, \ and\ \bibinfo {author} {\bibfnamefont
  {T.}~\bibnamefont {Minton}},\ }\bibfield  {title} {\enquote {\bibinfo {title}
  {{Qualification of Acktar Black Coatings for Space Application}},}\ }in\
  \href@noop {} {\emph {\bibinfo {booktitle} {Proceedings of the International
  Symposium on Materials in a Space Environment, Aix-En-Provence, France}}}\
  (\bibinfo {year} {2009})\BibitemShut {NoStop}%
\bibitem [{\citenamefont {Jhabvala}\ \emph {et~al.}(2016)\citenamefont
  {Jhabvala}, \citenamefont {Benford}, \citenamefont {Brekosky}, \citenamefont
  {Costen}, \citenamefont {Datesman}, \citenamefont {Hilton}, \citenamefont
  {Irwin}, \citenamefont {Maher}, \citenamefont {Manos}, \citenamefont
  {Miller}, \citenamefont {Moseley}, \citenamefont {Sharp}, \citenamefont
  {Staguhn}, \citenamefont {Wang},\ and\ \citenamefont
  {Wollack}}]{Jhabvala2016}%
  \BibitemOpen
  \bibfield  {author} {\bibinfo {author} {\bibfnamefont {C.~A.}\ \bibnamefont
  {Jhabvala}}, \bibinfo {author} {\bibfnamefont {D.~J.}\ \bibnamefont
  {Benford}}, \bibinfo {author} {\bibfnamefont {R.~P.}\ \bibnamefont
  {Brekosky}}, \bibinfo {author} {\bibfnamefont {N.~P.}\ \bibnamefont
  {Costen}}, \bibinfo {author} {\bibfnamefont {A.~M.}\ \bibnamefont
  {Datesman}}, \bibinfo {author} {\bibfnamefont {G.~C.}\ \bibnamefont
  {Hilton}}, \bibinfo {author} {\bibfnamefont {K.~D.}\ \bibnamefont {Irwin}},
  \bibinfo {author} {\bibfnamefont {S.~F.}\ \bibnamefont {Maher}}, \bibinfo
  {author} {\bibfnamefont {G.}~\bibnamefont {Manos}}, \bibinfo {author}
  {\bibfnamefont {T.~M.}\ \bibnamefont {Miller}}, \bibinfo {author}
  {\bibfnamefont {S.~H.}\ \bibnamefont {Moseley}}, \bibinfo {author}
  {\bibfnamefont {E.~H.}\ \bibnamefont {Sharp}}, \bibinfo {author}
  {\bibfnamefont {J.~G.}\ \bibnamefont {Staguhn}}, \bibinfo {author}
  {\bibfnamefont {F.}~\bibnamefont {Wang}}, \ and\ \bibinfo {author}
  {\bibfnamefont {E.~J.}\ \bibnamefont {Wollack}},\ }\bibfield  {title}
  {\enquote {\bibinfo {title} {Superconducting pathways through kilopixel
  backshort--under--grid arrays},}\ }\href {\doibase 10.1007/s10909-016-1487-y}
  {\bibfield  {journal} {\bibinfo  {journal} {Journal of Low Temperature
  Physics}\ ,\ \bibinfo {pages} {1--6}} (\bibinfo {year} {2016})}\BibitemShut
  {NoStop}%
\bibitem [{\citenamefont {Staguhn}\ \emph {et~al.}(2016)\citenamefont
  {Staguhn}, \citenamefont {Benford}, \citenamefont {Dowell}, \citenamefont
  {Fixsen}, \citenamefont {Hilton}, \citenamefont {Irwin}, \citenamefont
  {Jhabvala}, \citenamefont {Maher}, \citenamefont {Miller}, \citenamefont
  {Moseley}, \citenamefont {Sharp}, \citenamefont {Runyan},\ and\ \citenamefont
  {Wollack}}]{Staguhn2016}%
  \BibitemOpen
  \bibfield  {author} {\bibinfo {author} {\bibfnamefont {J.~G.}\ \bibnamefont
  {Staguhn}}, \bibinfo {author} {\bibfnamefont {D.~J.}\ \bibnamefont
  {Benford}}, \bibinfo {author} {\bibfnamefont {C.~D.}\ \bibnamefont {Dowell}},
  \bibinfo {author} {\bibfnamefont {D.~J.}\ \bibnamefont {Fixsen}}, \bibinfo
  {author} {\bibfnamefont {G.~C.}\ \bibnamefont {Hilton}}, \bibinfo {author}
  {\bibfnamefont {K.~D.}\ \bibnamefont {Irwin}}, \bibinfo {author}
  {\bibfnamefont {C.~A.}\ \bibnamefont {Jhabvala}}, \bibinfo {author}
  {\bibfnamefont {S.~F.}\ \bibnamefont {Maher}}, \bibinfo {author}
  {\bibfnamefont {T.~M.}\ \bibnamefont {Miller}}, \bibinfo {author}
  {\bibfnamefont {S.~H.}\ \bibnamefont {Moseley}}, \bibinfo {author}
  {\bibfnamefont {E.~H.}\ \bibnamefont {Sharp}}, \bibinfo {author}
  {\bibfnamefont {M.~C.}\ \bibnamefont {Runyan}}, \ and\ \bibinfo {author}
  {\bibfnamefont {E.~J.}\ \bibnamefont {Wollack}},\ }\bibfield  {title}
  {\enquote {\bibinfo {title} {Performance of backshort-under-grid kilopixel
  tes arrays for hawc+},}\ }\href {\doibase 10.1007/s10909-016-1509-9}
  {\bibfield  {journal} {\bibinfo  {journal} {Journal of Low Temperature
  Physics}\ ,\ \bibinfo {pages} {1--5}} (\bibinfo {year} {2016})}\BibitemShut
  {NoStop}%
\end{thebibliography}
\end{document}